\pdfoutput=1

\documentclass[aps,twocolumn,amsmath,amssymb,preprintnumbers,floatfix,prb,superscriptaddress,longbibliography]{revtex4-2}

\usepackage[utf8]{inputenc}
\usepackage{newtxtext}
\usepackage[upint]{newtxmath}
\usepackage{microtype}
\usepackage{textcomp}
\usepackage{eucal}
\usepackage{bm}
\usepackage{siunitx}
\usepackage{comment}
\usepackage{lipsum}

\usepackage{enumerate}
\usepackage{amsfonts}
\usepackage{amsmath}
\usepackage{amssymb}
\usepackage{color}
\usepackage{soul}

\usepackage{graphicx}

\usepackage[colorlinks,allcolors=blue]{hyperref}
\usepackage[capitalize]{cleveref}

\definecolor{DarkRed}{rgb}{0.65,0,0}%
\definecolor{Green}{rgb}{0,0.3,0.3}
\definecolor{Purple}{rgb}{0.3,0,0.65}
\definecolor{Red}{rgb}{1,0,0}
\definecolor{Blue}{rgb}{0,0,0.85}
\definecolor{Magenta}{rgb}{1,0,1}

\newcommand{\Imag}{{\Im\mathrm{m}}}   
\newcommand{\Real}{{\mathrm{Re}}}   
\newcommand{\im}{\mathrm{i}}        
\newcommand{\ve}[1]{\boldsymbol{#1}}

\newcommand{\vk}{{\ve{k}}} 

\newcommand{\e}[1]{\mathrm{e}^{#1}}

\newcommand{\veck}{\ve{k}}

\def\i{\mathrm{i}}

\newcommand{\g}{\underline{\gamma}}

\newcommand{\be}{\begin{equation}}
\newcommand{\ee}{\end{equation}}

\newcommand{\prlsection}[1]{\textit{#1}.\kern0.05em---\kern0.05em\ignorespaces}

\begin{document}
\title{Andreev reflection in altermagnets}
\author{Chi Sun}
\affiliation{Center for Quantum Spintronics, Department of Physics, Norwegian \\ University of Science and Technology, NO-7491 Trondheim, Norway}
\author{Arne Brataas}
\affiliation{Center for Quantum Spintronics, Department of Physics, Norwegian \\ University of Science and Technology, NO-7491 Trondheim, Norway}
\author{Jacob Linder}
\affiliation{Center for Quantum Spintronics, Department of Physics, Norwegian \\ University of Science and Technology, NO-7491 Trondheim, Norway}

\begin{abstract}
Recent works have predicted
materials featuring bands with a large spin-splitting distinct from ferromagnetic
and relativistically spin-orbit coupled systems. 
Materials displaying this property are known as altermagnets and feature a spin-polarized band structure reminiscent of a $d$-wave superconducting order parameter.
We here consider the contact between an altermagnet and a superconductor and determine how the altermagnetism affects the fundamental process of Andreev reflection. We show that the resulting charge conductance depends strongly on the interfacial orientation of the altermagnet relative to the superconductor, displaying features similar to normal metals or ferromagnets. The zero-bias conductance peaks present at the interface in the $d$-wave case are robust toward the presence of an altermagnetic interaction. 
Moreover, the spin conductance strongly depends on the orientation of the altermagnet relative the interface.
These results show how the 
anisotropic altermagnetic state can be probed by conductance spectroscopy and how it offers voltage control over
charge and spin currents that are 
modulated due to superconductivity.

\end{abstract}
\maketitle

\section{Introduction}
The interaction between magnetism and superconductivity is a major research topic in modern condensed matter physics \cite{buzdin_rmp_05, bergeret_rmp_05, linder_nphys_15, eschrig_rpp_15}. Its allure stems both from a fundamental viewpoint and cryogenic technology applications such as extremely sensitive detectors of radiation and heat as well as circuit components such as qubits and dissipationless diodes \cite{bergeret_rmp_18, amundsen_rmp_22}. 

To understand the transport of charge, spin, and heat in such structures, it is crucial to understand the basic transport mechanism involving the Cooper pair condensate: Andreev reflection \cite{andreev}. Whereas Andreev reflection in ferromagnetic materials has been studied in great detail \cite{eschrig_rpp_15}, antiferromagnetic materials have
received less attention. A particularly interesting example is recently discovered antiferromagnets \cite{ahn_prb_19, hayami_jpsj_19} that break time-reversal symmetry and feature a spin-splitting that does not originate from relativistic effects such as spin-orbit coupling \cite{pekar_zetf_64}. Dubbed altermagnets \cite{smejkal_prx_perspective_22} in the literature, these are spin-compensated magnetic systems with a huge momentum-dependent spin splitting even in collinearly ordered antiferromagnets. \textit{Ab initio} calculations have identified several possible material candidates that can host an altermagnetic state, including metals like RuO$_2$ and Mn$_5$Si$_3$ as well as semiconductors/insulators like MnF$_2$ and La$_2$CuO$_4$ \cite{yuan_prb_20, lopez-moreno_pccp_16, smejkal_sciadv_20, reichlova_arxiv_20, smejkal_prx_22}.

\begin{figure}[t!]
\includegraphics[width=0.87\columnwidth]{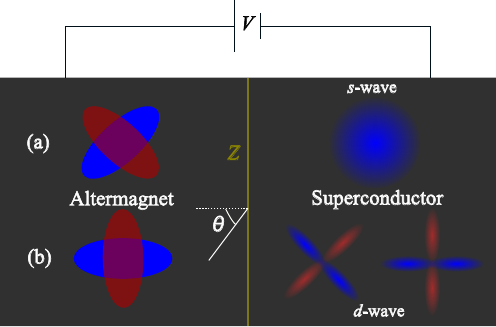}
	\caption{Andreev reflection is probed in a bilayer consisting of an altermagnet (AM) and a superconductor (SC). The order parameter in the SC can have a $s$-wave or $d$-wave symmetry, including different nodal orientations of the $d$-wave case. Different interface orientations are also considered, effectively rotating the spin-resolved Fermi surface in the AM for majority (blue ellipse) and minority (red ellipse) spin carriers. A voltage $V$ is applied to the system and the differential conductance provides information about the Andreev reflection.
	}
	\label{fig:model}
\end{figure}

The interaction between superconductivity and altermagnetism has only very recently started to be explored \cite{mazin_arxiv_22, ouassou_arxiv_23, zhang_arxiv_23, papaj_arxiv_23}. 
An interesting analogy exists between altermagnets and unconventional superconductivity in the high-$T_c$ cuprates where the order parameter has a $d$-wave symmetry in momentum space \cite{fischer_rmp_07}. Similarly, the band structure of altermagnets has a spin-resolved $d$-wave symmetry which mimics 
the structure of the $d$-wave superconducting order parameter (see Fig. \ref{fig:model}). 
Since hybrid structures of superconductors and magnetic materials are attracting wide interest due to their functional properties, we here consider Andreev reflection in an altermagnet (AM)/superconductor (SC) bilayer. We allow for both conventional $s$-wave superconductivity and unconventional $d$-wave superconductivity. Importantly, we allow for different crystallographic orientations of the interface between the materials to explore both how the nodal orientation of the SC order parameter and the spin-resolved Fermi surface orientation in the AM affect transport. 

We find that the altermagnetism strongly influences both charge and spin currents flowing into the SC for high-transparency contacts. Depending on the crystallographic orientation of the interface relative to the spin-polarized lobes of the altermagnetic Fermi surface, the zero-bias charge conductance peak in $d$-wave superconductors can be either enhanced or suppressed relative to the normal-state with increasing altermagnetic strength. Moreover, the spin conductance strongly depends on the orientation of the altermagnet relative the interface. Our findings demonstrate how the unique momentum-dependent spin polarization of the altermagnetic state is revealed in conductance spectroscopy by using superconductors.

\section{Theory}

The Hamiltonian for the AM, using a field operator basis $\psi = [\psi_\uparrow, \psi_\downarrow, \psi_\uparrow^\dag, \psi_\downarrow^\dag]^T$,  is given by
\begin{equation}
    \hat{H}_\text{AM}=
    \begin{pmatrix}
 \underline{H}_\text{AM}&0\\0&-\underline{H}_\text{AM}^*
\end{pmatrix},\; \underline{H}_\text{AM} =  -{\frac{\hbar^2\triangledown^2}{2 m_e}} - \mu + \alpha k_x k_y\underline{\sigma_z},
\end{equation}
in which $\alpha$ is the parameter that characterizes the altermagnetism strength, $\underline{{\sigma_z}}$ denotes the Pauli matrix, $m_e$ is the electron mass, $\mu$ is the chemical potential and $\underline{\ldots}$ is notation for a 2$\times$2 matrix. The four eigenpairs are obtained as: $E_1=E_+$ with $(1,0,0,0)^T$ for $e\uparrow$, $E_2=E_-$ with $(0,1,0,0)^T$ for $e\downarrow$, $E_3=-E_+$ with $(0,0,1,0)^T$ for $h\uparrow$ and $E_4=-E_-$ with $(0,0,0,1)^T$ for $h\downarrow$, using $e/h$ for electron/hole excitations. The eigenenergies are 
\begin{align}
E_\pm=\frac{\hbar^2(k_{x}^2+k_{y}^2)}{2m_e}-\mu \pm \alpha k_x k_y.
\end{align}

Considering an excitation with energy $E$, the $x$-components of the four possible wave vectors in the AM are given by $k_{e(h)\sigma,\pm} = \pm \hbar^{-1} \sqrt{2m_e(\mu \pm' E)  -\hbar^2k_y^2 + \alpha^2m_e^2k_y^2 \hbar^{-2}} - \alpha\sigma m_ek_y\hbar^{-2}$, where $\sigma=+1(-1)$ for spin-$\uparrow(\downarrow)$, $\pm'=+(-)$ for $e(h)$. The $\pm$ sign in the subscript denotes  propagation direction along $\pm x$. Translational invariance is assumed in the $y$-direction with associated
momentum $k_y$ of the incident particle. In the superconducting region, we use well-known expressions for the Hamiltonian and eigenenergies/states, allowing for both $s$-wave and $d$-wave symmetries  (see appendix for details).

The altermagnetic Hamiltonian modifies the standard expressions for the charge current and boundary conditions satisfied by the scattering wavefunctions. To see this, consider for concreteness an $e\uparrow$ incident from the AM side of an AM/SC bilayer. We  have
\begin{align}
\Psi_{\text{AM},e\uparrow} &=
\begin{pmatrix}
1\\0
\end{pmatrix}e^{ik_{e\uparrow,+}x}+r
\begin{pmatrix}
 1\\0   
\end{pmatrix}e^{ik_{e\uparrow,-}x} + r_A
\begin{pmatrix}
    0\\1
\end{pmatrix}e^{ik_{h\downarrow,+}x},\notag\\
\Psi_{\text{SC},e\uparrow} &=
t\begin{pmatrix}
u_+\\v_+ e^{-i\gamma_+}
\end{pmatrix}e^{iq_{e,+}x}+t_A
\begin{pmatrix}
 v_- e^{i \gamma_-}\\u_-  \end{pmatrix}e^{-iq_{h,-}x},
\label{eq:wavefunctions}
\end{align}
in which $r$, $r_A$, $t$ and $t_A$ describe the normal reflection, Andreev reflection, normal transmission, and Andreev transmission, respectively.
We consider a superconducting gap which can be anisotropic, $\Delta = \Delta_0g(\theta_S)$ with $g(\theta_S)=1$ for the $s$-wave case and $g(\theta_S) = \cos{(2\theta_S-2\beta)}$ for the $d$-wave case, where $\e{i\gamma_\pm} = g(\theta_\pm)/|g(\theta_\pm)|$ with $\theta_+ = \theta_S$ and $\theta_- = \pi-\theta_S$ are defined. The scattering angle $\theta_S$ in the SC is determined from $\theta$ in the AM by using conservation of momentum $k_y.$

To derive the boundary condition for the $e\uparrow$ incident, antisymmetrization of the altermagnetic term 
\begin{align}
\alpha k_xk_y\underline{\sigma_z} \to \frac{\alpha k_y}{2} \{k_x,\Theta(-x)\} \underline{\sigma_z}
\end{align}
    is necessary to ensure hermiticity of the Hamilton-operator, where $\Theta(x)$ is the step function. Above, $k_x = -\i\partial_x$. Applying $H\Psi = E \Psi$ and integrating over $[-\epsilon,\epsilon]$ with $\epsilon \rightarrow 0$, we obtain $\Psi_{\text{AM},e\uparrow}\big|_{x=0}=\Psi_{\text{SC},e\uparrow}\big|_{x=0}=(f,g)^T$ and
\begin{align}
\partial_x\Psi_{\text{SC},e\uparrow}\big|_{x=0}-\partial_x\Psi_{\text{AM},e\uparrow}\big|_{x=0}=\begin{pmatrix}
k_{\alpha,+1}f\\k_{\alpha,-1}g
\end{pmatrix},
\end{align} 
where 
\begin{align}
k_{\alpha,\sigma} =\frac{2m_e}{\hbar^2}(U_0 + \frac{\i\alpha k_y\sigma}{2}).
\label{eq:k_alpha_sigma}
\end{align}
Here the imaginary number $\im$ appears in $k_{\alpha,\sigma}$ since we consider $k_y$ invariance (unlike $k_x = -\im\partial_x$).
The boundary conditions for incident $e\downarrow$, $h\uparrow$ and $h\downarrow$ particles can be found in the appendix.

To compute the conductance of the junction, the charge current produced by all possible types of incoming quasiparticles toward the interface must be considered
. The electric current is computed by taking the quantum mechanical expression for the charge current and multiplying it with the density of states (DOS) and distribution function of the incident particle. The DOS of quasiparticles in the superconducting region is well-known but is worth presenting in the altermagnetic region. We consider again an incident $e\uparrow$ with energy $E$ from the AM side for concreteness. We have $E = E_+ =\frac{\hbar^2(k_{x}^2+k_{y}^2)}{2m_e}-\mu + \alpha k_x k_y.$
The general expression for 2D DOS of a band $E(\veck)$ is given by
\begin{equation}
   N(E) = \frac{1}{4\pi^2} \int \frac{dl}{|\nabla_{\veck} E(\veck)|}, 
\label{eq:def1_dos}
\end{equation}
which can be used to compute the $\vk$-anisotropic DOS in the altermagnetic case. When $\alpha < \hbar^2/m_e \equiv \alpha_c$, a constant energy contour defines an elliptical energy surface in $\vk$-space for $e\uparrow$. The ellipse has semi-major (minor) axis $a$ ($b$), which can be obtained as 
\begin{align}
a = \sqrt{\frac{2m_e(\mu+E)}{\hbar^2-m_e \alpha}},\; b = \sqrt{\frac{2m_e(\mu+E)}{\hbar^2+m_e\alpha}}.
\end{align}
On the other hand, when $\alpha > \alpha_c$, the energy dispersion corresponds to a hyperbola, which can not define a closed integral path. Therefore, we confine our attention to $\alpha < \alpha_c$ in this work. 


The quantum mechanical charge current density for $e\uparrow$ channel in the AM is given by 
\begin{equation}
 j_{Q,e\uparrow} = -\frac{e\hbar}{m_e}[\Imag{\{f^*\nabla f \}} + \Imag{\{g^*\nabla g \}}] - \frac{e\alpha k_y}{\hbar}(|f|^2-|g|^2).
 \label{eq:j_Q1,eup}
\end{equation}
We can compute the total charge current flowing in the AM by using Eq. (\ref{eq:j_Q1,eup}) in the $e\uparrow$ channel and integrating over all incoming modes after multiplying $j_{Q,e\uparrow}$ with the distribution function for the incoming particles. 
Assume that a voltage is applied across the AM/SC junction so that the distribution function for electrons [holes] is $f(E-eV)$ [$f(E+eV)$] on the AM side while it is $f(E)$ for quasiparticles on the SC side. For instance, an incoming hole from the AM side can be Andreev-reflected into the $e\uparrow$ channel as  $\psi = r_{A}\begin{pmatrix}
1\\0
\end{pmatrix}e^{ik_{e\uparrow,-}x}$ and contributes with a current 
\begin{align}
j_{Q,e\uparrow} =-e|r_{A}|^2(\frac{\hbar k_{e\uparrow,-}}{m_e}+\frac{\alpha k_y}{\hbar}).
\end{align}
The total charge current $I$ is then obtained by first computing the total electric current flowing in the $e$ and $h$ channels for both spins on the AM side, each contribution determined by $j_{Q,i} f_i(E)$ where $j_{Q,i}$ is the charge current density produced by an incoming particle channel $i$, $f_i(E)$ is the distribution function for channel $i$, and then integrating over all energies and all possible transverse modes via $\int dk_x = \int dE (dk_x/dE)$ and $\int dk_y$. In performing the integration over transverse modes, conservation of momentum $k_y$ needs to be taken into account. We find that the charge current carried in the $e$ and $h$ channels are equal, and it thus suffices to consider only transport in the $e$ channel for spin-$\uparrow$ and $\downarrow$. Specifically, the total charge current $I_\sigma$ flowing in the electron channel for spin $\sigma$ is obtained as:
\begin{align}\label{eq:current}
I_\sigma \propto \int \int dE \frac{dk_x}{dE} d k_y j_{Q,i}^{e,\sigma}(E,\theta)f_i(E) 
\end{align}
where $j_{Q,i}^{e,\sigma}$ is the charge current produced in the $e,\sigma$ channel from a particle that is incoming from channel $i$, giving a total current $I = I_\uparrow+I_\downarrow$.
Andreev-reflection and normal-reflection contribute to the conductance qualitatively in the same way as in the BTK model (the former enhancing and the latter suppressing the conductance) \cite{btk}.
The conductance is then $G(V) = dI/dV$ and we normalize it against the high-voltage conductance (normal-state) $G_0 = \lim_{eV \gg \Delta_0} G(V)$, which cancels the proportionality constant, which includes the area of the junction, in Eq. (\ref{eq:current}). The spin current $I_S$ is obtained by computing the difference $I_\uparrow-I_\downarrow$ between the currents carried by $e\uparrow$ and $e\downarrow$, and the spin conductance is $G_S = dI_S/dV$ with a similar normalization as for the charge current.

We will show how the conductance of the AM/SC junction depends strongly on the crystallographic orientation of the interface between the materials. This can be modeled by replacing $\alpha k_xk_y \to \alpha(k_x^2-k_y^2)/2$ in $\underline{H}_\text{AM}$, corresponding to a $45$ degree rotation of the interface. This leads to different expressions for the wavevectors 
\begin{align}\label{eq:wavevectorkx2ky2}
k_{e(h)\sigma,\pm} = \pm \sqrt{\frac{2m_e(\mu \pm' E + \sigma\alpha k_y^2/2) -\hbar^2k_y^2}{\hbar^2 + m_e\sigma\alpha}},
\end{align}
boundary condition
\begin{align}
\partial_x\Psi_{\text{SC},e\uparrow}|_{x=0} - \begin{pmatrix}
1 + m_e\alpha\hbar^{-2} \partial_x f \\
1 - m_e\alpha\hbar^{-2} \partial_x g\\
    \end{pmatrix} = \frac{2mU_0}{\hbar^2}\begin{pmatrix}
 f \\g\\
    \end{pmatrix},
    \end{align}
and the charge current density
\begin{align}
j_{Q, e(h)\sigma} = \text{Im}\{f^*\nabla f\}(-e\hbar/m_e \mp' e\alpha\sigma/\hbar) \notag\\
+ \text{Im}\{g^*\nabla g\}(-e\hbar/m_e \pm' e\alpha\sigma/\hbar).
\end{align}
The boundary conditions for incident $e\downarrow$, $h\uparrow$ and $h\downarrow$ particles can be found in the appendix. A similar procedure as described earlier can then be used to compute the charge and spin conductances of the junction.

\begin{figure}[t!]
\includegraphics[width=0.99\columnwidth]{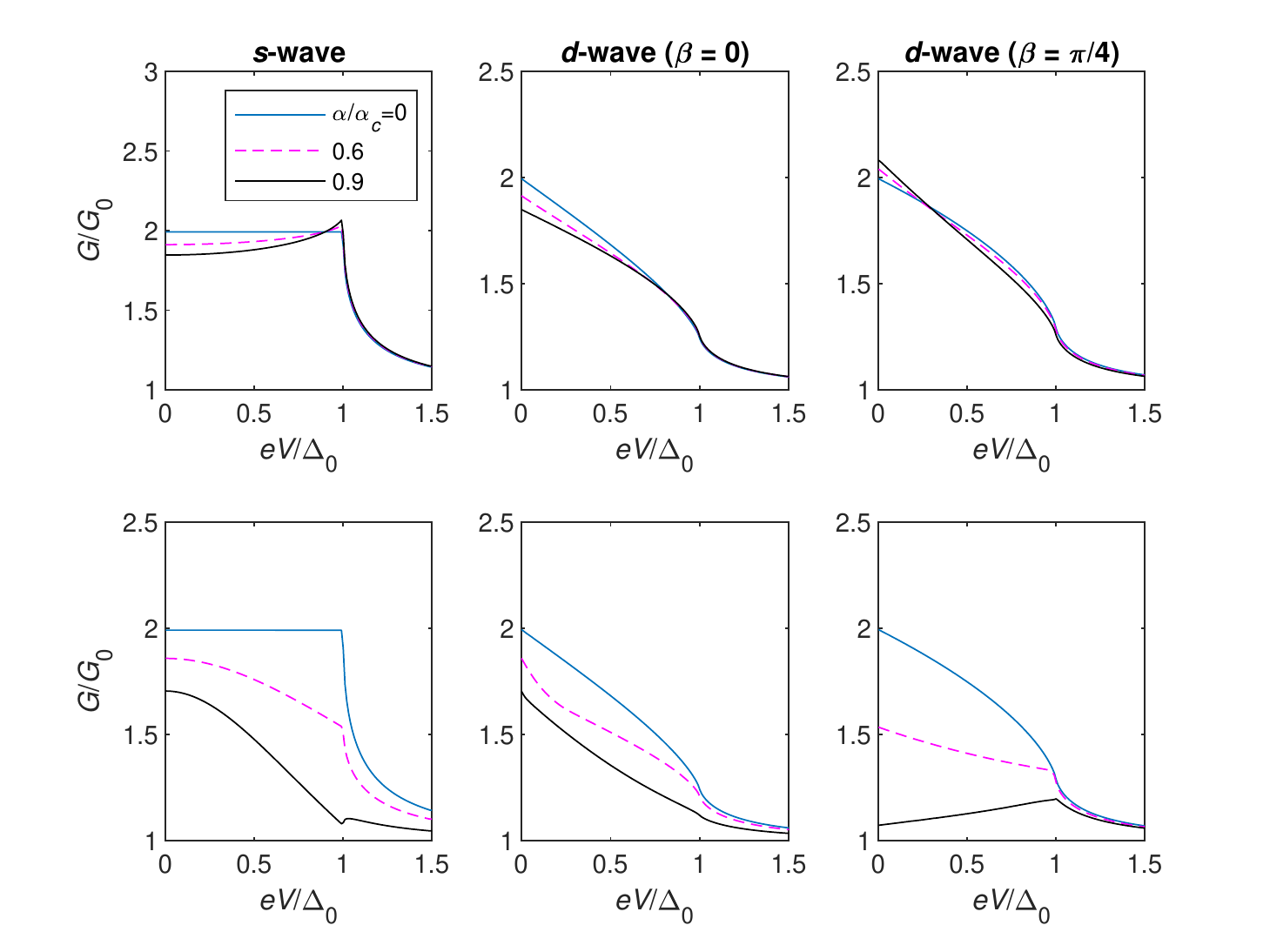}
	\caption{(Color online) Normalized charge conductance $G/G_0$ as a function of bias voltage for different types of AM/SC junctions. The barrier is set to $Z=0$, which describes a high-transparency contact. The columns correspond to different superconducting order parameter symmetries. \textit{Upper row:} case (a) in Fig. \ref{fig:model} [AM term $\alpha k_xk_y$ in $\underline{H}_{\text{AM}}$]. \textit{Lower row:} case (b) in Fig. \ref{fig:model} [AM term $\alpha (k_x^2-k_y^2)/2$ in $\underline{H}_{\text{AM}}$].
	}
	\label{fig:Zzero}
\end{figure}

\section{Results}

The dimensionless parameter $Z= \frac{m_e U_0}{\hbar^2 k_F}$ with $k_F = \sqrt{\frac{2m_e\mu}{\hbar}}$ characterizes the quality of electric contact between the AM and SC \cite{btk}. The high-transparency limit $Z\ll 1$ is routinely achievable experimentally using point-contact spectroscopy measurements \cite{naidyuk_book_04, daghero_sst_10} or very high-quality interfaces. A tunneling interface, modeled by $Z=3$ in this work, can be achieved with the same experimental technique by increasing the tip-sample distance or by explicitly inserting a thin insulator between the AM and SC. Both transport regimes are interesting, and the altermagnetic interactions reveal themselves differently in these two cases. 

To understand the results for the conductance, it is useful to consider the wavevectors of the incident electrons with the corresponding Andreev-reflected holes. In the NM case, there is only a very slight mismatch between the wavevectors of the incident and Andreev-reflected particles as the sign of the energy changes: the wavevector is proportional to factor $\sqrt{\mu+E}$ vs. $\sqrt{\mu-E}$. However, for ferromagnetic materials (FMs), there is a much larger mismatch between these wavevectors due to the presence of a (momentum-independent) spin-splitting or exchange energy $J_\text{ex}$: $\sqrt{\mu + J_\text{ex}}$ vs. $\sqrt{\mu-J_\text{ex}}$. This large change in momentum suppresses the Andreev-reflection process as $J_\text{ex}$ increases.

We can now compare this with the altermagnetic case. For simplicity, let us focus on particles close to normal incidence, $k_y \to 0$, which contribute the most to the transport across the junction. In the $k_xk_y$ orientation of the spin-bands of the AM, the wavevectors of the incident and Andreev-reflected particles are then almost equal, distinguished only by their sign in energy, just like the NM case. In contrast, in the $k_x^2-k_y^2$ case, the wavevectors can be strongly mismatched even for $k_y\to 0$ as seen from Eq. (\ref{eq:wavevectorkx2ky2}). This is similar to the FM case.

With increasing $k_y$, however, the mismatch increases in the $k_xk_y$ case while it decreases in the $k_x^2-k_y^2$ case. This is different from both the NM and FM cases and is a unique feature of the altermagnetic band structure. For larger $k_y$, Andreev-reflection thus becomes less favorable in the $k_xk_y$ orientation compared to normal incidence, whereas the opposite is true in the $k_x^2-k_y^2$ orientation.

The conductance in the high-transparency case is shown in Fig. \ref{fig:Zzero}. In this case, increasing the magnitude of the spin-splitting $\alpha$ in the altermagnetic band structure substantially changes the conductance. In the $s$-wave and $d$-wave $\beta=0$ cases, both known 
not to feature interfacial bound-states, the conductance is suppressed with increasing $\alpha$. In the $d$-wave $\beta=\pi/4$ case, known to feature zero-energy bound states at interfaces and defects, the conductance is either enhanced or suppressed relative to the normal state depending on the orientation of the spin-polarized elliptical Fermi-surfaces of majority and minority spin carriers.

As explained before, it can be seen that the influence of the $k_x^2-k_y^2$ altermagnetism on the conductance for the orientation shown in case (b) in Fig. \ref{fig:model} (corresponding to the lower row of Fig. \ref{fig:Zzero}) is similar to that of a conventional FM/SC junction \cite{FMSC}: the magnetic interaction simply suppresses the conductance. This can be understood physically from the fact that the most dominant trajectories contributing to transport in the junction are the ones normal to the interface. For such directions of $\theta$, the band structure of the altermagnet is qualitatively similar to a ferromagnet in that one spin species dominates over the other independently of momentum. Our model can also be directly applied to describe the conventional FM by considering $\underline{H}_\text{FM} =  -{\frac{\hbar^2\triangledown^2}{2 m_e}} - \mu + J_\text{ex}\underline{\sigma_z}$,  which can give the same trends as shown in \cite{FMSC}. Interested readers are referred to the appendix for more details.

\begin{figure}[t!]
\includegraphics[width=0.99\columnwidth]{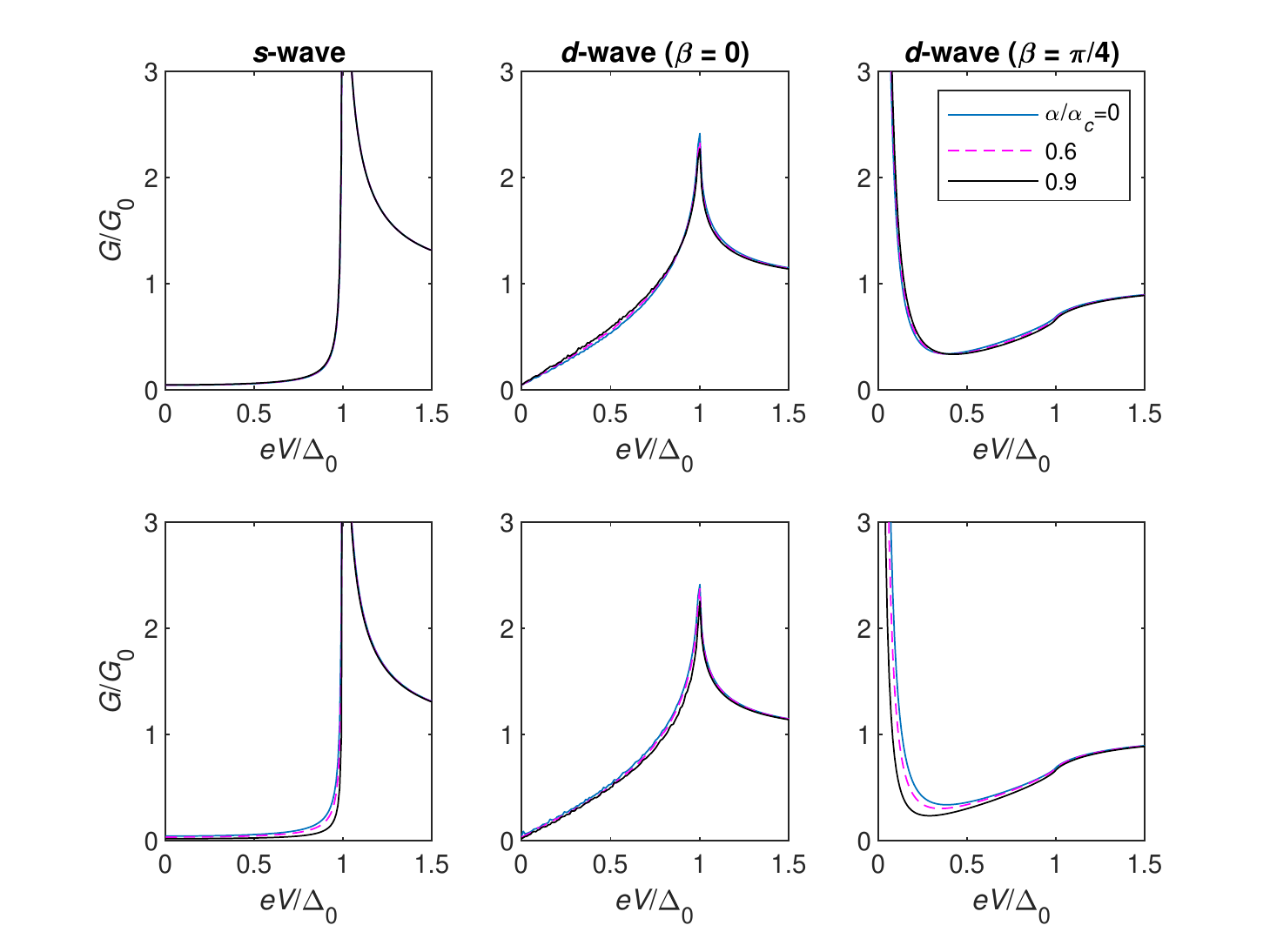}
	\caption{(Color online) Normalized charge conductance $G/G_0$ as a function of bias voltage for different types of AM/SC junctions. The barrier is set to $Z=3$, which describes a low-transparency contact. The columns correspond to different superconducting order parameter symmetries. \textit{Upper row:} case (a) in Fig. \ref{fig:model} [AM term $\alpha k_xk_y$ in $\underline{H}_{\text{AM}}$]. \textit{Lower row:} case (b) in Fig. \ref{fig:model} [AM term $\alpha (k_x^2-k_y^2)/2$ in $\underline{H}_{\text{AM}}$].
	}
	\label{fig:Zthree}
\end{figure}

Case (a) in Fig. \ref{fig:model} (corresponding to the upper row of Fig. \ref{fig:Zzero}) is more complex and interesting. In this $k_xk_y$ AM case, there is no spin-polarization for normal incidence $\theta=0$, while spin-$\downarrow$ is the majority carrier for incident electrons with $\theta>0$ and spin-$\uparrow$ is the majority carrier for $\theta<0$. The total spin polarization of the incident
particles cancel since the majority and minority spin bands contribute equally when integrating over all possible angles of incidence toward the AM/SC interface. Therefore the AM behaves similarly
to a
NM with zero spin-polarization, as mentioned before.  Compared with the FM, the reduction in spin-polarization for incident particles then causes a lesser suppression of the charge conductance, consistent with the upper row of Fig. \ref{fig:Zzero} 
 for $s$-wave and $d$-wave $\beta=0$. On the other hand, the conductance relative the normal state increases with altermagnetism for the $d$-wave $\beta=\pi/4$ SC, which corresponds to the behavior with a higher effective barrier introduced by altermagnetism, as will be explained below, based on comparison with the
the NM/$d$-wave SC $\beta=\pi/4$ shown in Fig. (2c) in \cite{tanaka_prl_95}.  Note that in the left top of Fig. \ref{fig:Zzero}, a slight peak at the gap edge appears for $\alpha/\alpha_c=0.9$, which is similar to the conductance behavior when adding a weak barrier (e.g., $Z=0.5$) at the interface of a NM/$s$-wave SC bilayer (e.g., Fig. 7 in Ref. \cite{btk}). Here this weak effective barrier is introduced by and proportional to the altermagnetism strength, i.e., the second term in $k_{\alpha,\sigma} =\frac{2m_e}{\hbar^2}(U_0 + \frac{\im\alpha k_y\sigma}{2}) $ in the boundary condition described by Eq. (\ref{eq:k_alpha_sigma}).  

In Fig. \ref{fig:Zthree}, we show the case of a tunneling interface between the AM and SC for completeness. 
In this case, the altermagnetism has less effect on the charge conductance, even for large values $\alpha/\alpha_c=0.9$. However, it is interesting to note that the zero-bias peak present for a $d$-wave $\beta=\pi/4$ order parameter \cite{hu_prl_94, tanaka_prl_95} survives for both orientations of the AM [case (a) and (b) in Fig. \ref{fig:model}]. This suggests that the  zero-bias conductance peaks known to be present at $d$-wave interfaces are robust toward the presence of altermagnetism. Furthermore, comparing Fig. \ref{fig:Zzero}, center top, to Fig. \ref{fig:Zthree}, center top, it can be seen that
going from the high transparency to the tunneling limit can reverse
the dependence of the conductance on the altermagnet strength relative to the normal state, which might be probed in experiments. This behavior can be understood by comparing with the NM/$d$-wave SC $\beta=0$ bilayer ( Fig. 2 in \cite{tanaka_prl_95}): In the high transparency limit with $Z=0$, the second term in $k_{\alpha,\sigma}$ acts as the only effective barrier whose strength increases with $\alpha$, playing a role as a weak $Z$. In the low transparency limit with $Z=3$,  the second term in $k_{\alpha,\sigma}$ can partially compensate $Z=3$ in the first term, giving rise a slightly lower but still strong $Z$.

\begin{figure}[t!]
\includegraphics[width=0.99\columnwidth]{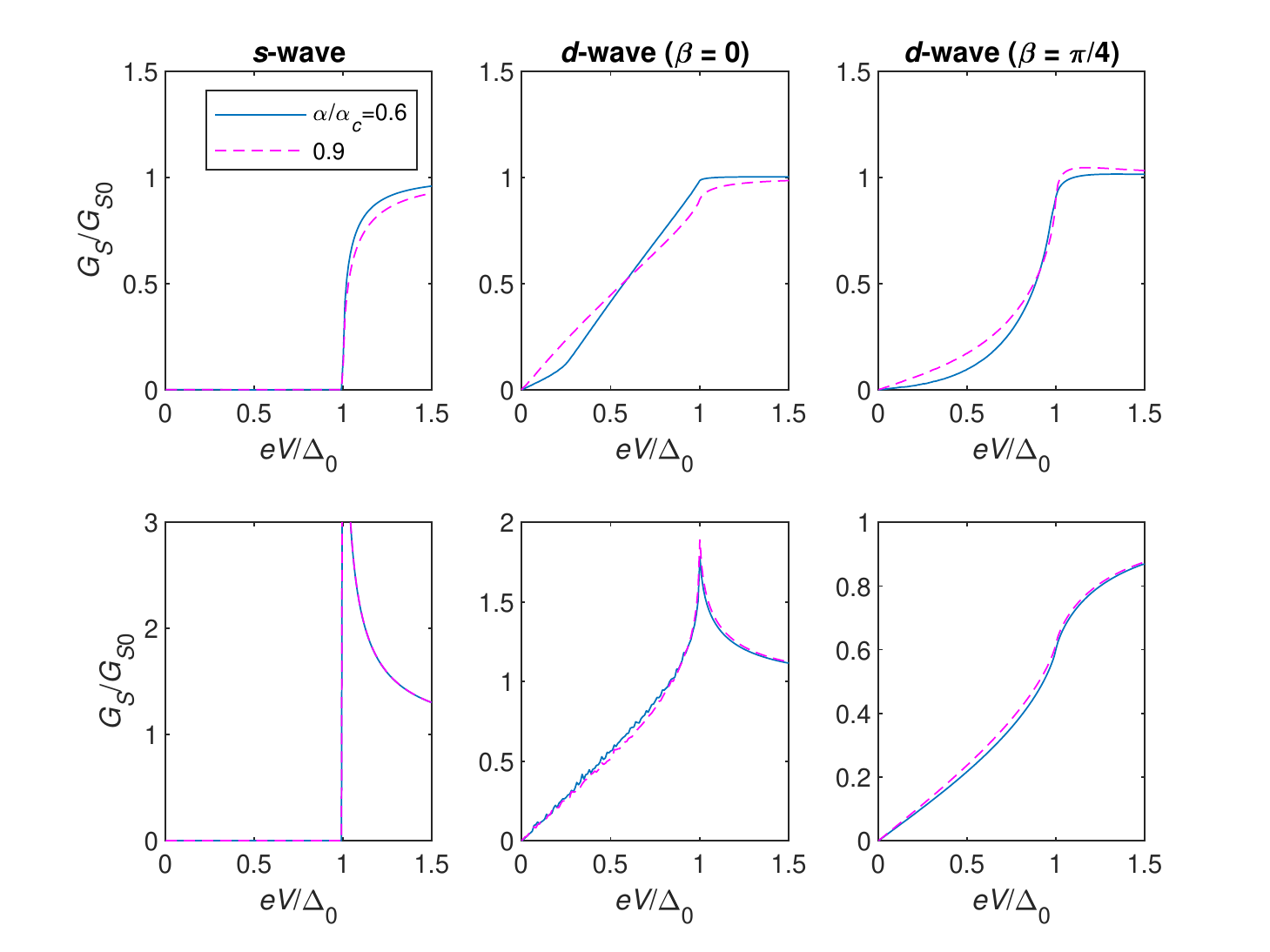}
	\caption{(Color online) Normalized spin conductance $G_S/G_{S0}$ as a function of bias voltage for case (b) in Fig. \ref{fig:model} [AM term $\alpha (k_x^2-k_y^2)/2$ in $\underline{H}_{\text{AM}}$]. The columns correspond to different superconducting order parameter symmetries. \textit{Upper row:} high-transparency contact $Z=0$. \textit{Lower row:} low-transparency contact $Z=3$.
	}
	\label{fig:Spin}
\end{figure}

Finally, we consider the spin-polarization properties of the current flowing in the junction. For case (a) in Fig. \ref{fig:model}, similar to NM, the spin conductance $G_S$ is zero since the total spin polarization of the incident particles 
cancels upon averaging over all incident angles.
For case (b) in Fig. \ref{fig:model}, the transmitted current is spin-polarized and shows similar behavior as a ferromagnet/superconductor bilayer \cite{FMSC}. The magnitude of the spin conductance $G_S$ vanishes for $\alpha=0$. The low-transparency case is considered in the lower row of Fig. \ref{fig:Spin}. Similarly to the charge conductance, the altermagnetic interaction has very little impact on the results in this case. Therefore, high-transparency contacts between altermagnets and superconductors will offer the clearest transport signature of the altermagnetic interaction.

In this work, two representative AMs with 0 and 45-degree rotation relative to the interface are investigated. As for the AM with arbitrary rotation, it can be modeled based on the combination of our established 0 and 45 degree cases, i.e., 
\begin{equation}
    \underline{H}_\text{AM} =  -{\frac{\hbar^2\triangledown^2}{2 m_e}} - \mu + \alpha_1 k_x k_y\underline{\sigma_z} + \alpha_2 (k_x^2- k_y^2)\underline{\sigma_z}/2,
\end{equation}
in which two different altermagnetism strength parameters $\alpha_1$ and $\alpha_2$ are introduced and the arbitrary angle is determined by $\theta_\alpha = \frac{1}{2}\arctan(\alpha_1/\alpha_2)$. More details can be found in the appendix. 

We also comment on the altermagnetism ratios used in the plots. We have defined the altermagnetism strength as $\alpha=n_\alpha\alpha_c$, in which $\alpha_c=\hbar^2/m_e$ is its critical value and $n_\alpha$ is the ratio, e.g., $(0,0.6,0.9)$. In terms of energy, the ratio between the altermagnetic and kinetic coefficients are $\alpha/\frac{\hbar^2}{2m_e}=2n_\alpha$. Previous \textit{ab initio} calculations have predicted spin splittings of order eV for metallic altermagnets. If we assume that the Fermi energy $\mu = \hbar^2k_F^2/2m_e$ in the normal state is of order eV, then we note that our choice of $n_\alpha=0.6$ corresponds to a maximal altermagnetic spin splitting in $\boldsymbol{k}$-space (roughly approximated as $\alpha k_F^2$) of similar magnitude as the \textit{ab initio} calculations.

\section{Summary} 
In conclusion, we have shown that charge and spin conductances are strongly affected by altermagnetism for junctions with high-quality interfaces. The zero-bias conductance peaks present for $d$-wave superconductors remain robust in the presence of altermagnetism
. The spin conductance demonstrates a strong dependence on the orientation of the altermagnetic crystal structure relative the interface. Our predicted effects can be tested experimentally using 
a metallic altermagnet such as RuO$_2$, and point the way toward a further investigation of interesting spintronics effects in heterostructures comprised of altermagnets and superconductors.

\acknowledgments

S. Rex, J. Danon, A. Qaiumzadeh, and J. A. Ouassou are thanked for useful discussions. 
This work was supported by the Research
Council of Norway through Grant No. 323766 and its Centres
of Excellence funding scheme Grant No. 262633 “QuSpin.” Support from
Sigma2 - the National Infrastructure for High Performance
Computing and Data Storage in Norway, project NN9577K, is acknowledged.

\appendix

\begin{widetext}

\section{Wave vectors in the AM}
The Hamiltonian for the altermagnet (AM), using a field operator basis $\psi = [\psi_\uparrow, \psi_\downarrow, \psi_\uparrow^\dag, \psi_\downarrow^\dag]^T$, is given by
\begin{equation}
    \hat{H}_\text{AM}=
    \begin{pmatrix}
 \underline{H}_\text{AM}&0\\0&-\underline{H}_\text{AM}^*
\end{pmatrix}
\end{equation}
with
\begin{equation}
\underline{H}_\text{AM} =  -{\frac{\hbar^2\triangledown^2}{2 m_e}} - \mu + \alpha\underline{\sigma_z}k_x k_y,
\end{equation}
 in which $\alpha$ is the parameter that characterizes the altermagnetism strength, $\underline{{\sigma_z}}$ denotes the Pauli matrix, $m_e$ is the electron mass, $\mu$ is the chemical potential and $\underline{\ldots}$ is notation for a 2$\times$2 matrix. The four eigenpairs are obtained as: $E_1=E_+$ with $(1,0,0,0)^T$ for $e\uparrow$, $E_2=E_-$ with $(0,1,0,0)^T$ for $e\downarrow$, $E_3=-E_+$ with $(0,0,1,0)^T$ for $h\uparrow$ and $E_4=-E_-$ with $(0,0,0,1)^T$ for $h\downarrow$. The eigenenergies are described by  
\begin{equation}
E_\pm=\frac{\hbar^2(k_{x}^2+k_{y}^2)}{2m_e}-\mu \pm \alpha k_x k_y.
\label{eq:Epm}
\end{equation}

Applying $E_1=E_2=E_3=E_4=E$, the $x$-components of the wave vectors in the AM are given by
\begin{equation}
   k_{e\uparrow,\pm}=\pm\frac{1}{\hbar}\sqrt{2m_e(\mu+E)-\hbar^2 k_y^2 +\frac{\alpha^2 m_e^2 k_y^2}{\hbar^2}}-\frac{\alpha m_e k_y}{\hbar^2}, 
\label{eq:keup}
\end{equation}
\begin{equation}
k_{e\downarrow,\pm}=\pm\frac{1}{\hbar}\sqrt{2m_e(\mu+E)-\hbar^2 k_y^2 +\frac{\alpha^2 m_e^2 k_y^2}{\hbar^2}}+\frac{\alpha m_e k_y}{\hbar^2}, 
\label{eq:kedn}
\end{equation}
\begin{equation}
   k_{h\uparrow,\pm}=\pm\frac{1}{\hbar}\sqrt{2m_e(\mu-E)-\hbar^2 k_y^2 +\frac{\alpha^2 m_e^2 k_y^2}{\hbar^2}}-\frac{\alpha m_e k_y}{\hbar^2},
\label{eq:khup}
\end{equation}
\begin{equation}
   k_{h\downarrow,\pm}=\pm\frac{1}{\hbar}\sqrt{2m_e(\mu-E)-\hbar^2 k_y^2 +\frac{\alpha^2 m_e^2 k_y^2}{\hbar^2}}+\frac{\alpha m_e k_y}{\hbar^2},
\label{eq:khdn}
\end{equation}
 in which the $\pm$ sign in the subscript denotes the propagation direction along the $\pm x$. Here we assume translational invariance in the $y$-direction with 
an associated conserved momentum $k_y$. The momentum $k_y$ of the incident particle appearing in Eqs. (\ref{eq:keup}-\ref{eq:khdn}) is determined by the Fermi surface of the incident particle, which is described as follows.

Consider an $e\uparrow$ particle in the AM. We then have $E = E_+ =\frac{\hbar^2(k_{x}^2+k_{y}^2)}{2m_e}-\mu + \alpha k_x k_y$ in Eq. (\ref{eq:Epm}), which defines an elliptical Fermi surface in the $\boldsymbol{k}$-space when $\alpha<\hbar^2/m_e\equiv\alpha_c$.
On the other hand, Eq. (\ref{eq:Epm}) corresponds to a hyperbola when $\alpha > \alpha_c$, which can not define a closed integral path. Therefore, we confine $\alpha < \alpha_c$ in this work. The general equation of the ellipse is given by
\begin{align}
\frac{\hbar^2k_{x}^2}{2m_e}+\alpha k_x k_y + \frac{\hbar^2 k_{y}^2}{2m_e}  - (\mu + E) = 0,
\end{align}
from which the semi-major (minor) axis can be obtained as
\begin{align}
    a_1 &= \sqrt{\frac{2m_e(\mu+E)}{\hbar^2-m_e \alpha}},\quad
    b_1 &= \sqrt{\frac{2m_e(\mu+E)}{\hbar^2+m_e\alpha}}.
\end{align}
Consequently, the wave vectors on the Fermi surface of $e\uparrow$ in the AM are described by
\begin{align}
k_{y,e\uparrow} &= r_1 \sin\theta
,\quad k_{x,e\uparrow} = r_1 \cos\theta,\quad r_1 &= \frac{a_1 b_1}{\sqrt{b_1^2 \cos^2(\theta+\pi/4)+a_1^2 \sin^2(\theta+\pi/4)}},
\label{eq:ky_eup}
\end{align}
in which $\theta$ is the incident angle in the AM with respect to the $x$-axis. Therefore, we use $k_y = k_{y,e\uparrow}$ in Eq. (\ref{eq:keup}-\ref{eq:khdn}) to get the $x$-component of the wave vector belonging to incident $e\uparrow$ particles on the AM side.

Similarly, we can obtain the wave vectors on the Fermi surface of $e\downarrow$, $h\uparrow$ and $h\downarrow$ particles in the AM, i.e.,
\begin{align}
k_{y,e\downarrow} &= r_2 \sin\theta
,\quad k_{x,e\downarrow} = r_2 \cos\theta,\notag \\ r_2 &= \frac{a_2 b_2}{\sqrt{b_2^2 \cos^2(\theta-\pi/4)+a_2^2 \sin^2(\theta-\pi/4)}},\quad   a_2 = \sqrt{\frac{2m_e(\mu+E)}{\hbar^2-m_e \alpha}},\quad  b_2= \sqrt{\frac{2m_e(\mu+E)}{\hbar^2+m_e\alpha}} \label{eq:ky_edn}\\
k_{y,h\uparrow} &= r_3 \sin\theta
,\quad k_{x,h\uparrow} = r_3 \cos\theta
, \notag \\ r_3 &= \frac{a_3 b_3}{\sqrt{b_3^2 \cos^2(\theta+\pi/4)+a_3^2 \sin^2(\theta+\pi/4)}},\quad a_3 = \sqrt{\frac{2m_e(\mu-E)}{\hbar^2-m_e \alpha}}, \quad b_3= \sqrt{\frac{2m_e(\mu-E)}{\hbar^2+m_e\alpha}} \label{eq:ky_hup}\\k_{y,h\downarrow} &= r_4 \sin\theta
,\quad k_{x,h\downarrow} = r_4 \cos\theta
, \notag \\r_4 &= \frac{a_4 b_4}{\sqrt{b_4^2 \cos^2(\theta-\pi/4)+a_4^2 \sin^2(\theta-\pi/4)}},\quad  a_4 = \sqrt{\frac{2m_e(\mu-E)}{\hbar^2-m_e \alpha}}, \quad b_4= \sqrt{\frac{2m_e(\mu-E)}{\hbar^2+m_e\alpha}}.
\label{eq:ky_hdn}
\end{align}
By inserting $k_y=k_{y,e\downarrow}$, $k_y=k_{y,h\uparrow}$ and $k_y=k_{y,h\downarrow}$ into Eq. (\ref{eq:keup}-\ref{eq:khdn}) we can get the  $x$-components of the wave vectors induced by $e\downarrow$, $h\uparrow$ and $h\downarrow $ incidents on the AM side, respectively, which will appear in the wave functions to describe the propagation along the $x$-direction.

Note the relation between the two $x$-components of the wave vectors involved, e.g., $k_{x,e\uparrow}$ and $k_{e\uparrow,\pm}$: $k_{x,e\uparrow}$ is the $x$-component of the wave vector of $e\uparrow$ particle on the Fermi surface for a given value of the angle $\theta$, and is thus uniquely defined. Instead, $k_{e\uparrow,\pm}$ are the two possible solutions for the $x$-component of the momentum on the Fermi surface which both have the same value for $k_y$. Thus, $k_{e\uparrow,\pm}$ can be used to describe the $x$-component of incident and reflected $e\uparrow$ particles for a given $k_y$-value. Only when considering the $e\uparrow$ incident from the AM with $k_y=k_{y,e\uparrow}$, $k_{x,e\uparrow}$ is equivalent to either $k_{e\uparrow,+}$ or $k_{e\uparrow,-}$ depending on the value of $\theta$. To construct the wave functions, we have thus assumed $k_y$ invariance and include the $x$-components of the wave vectors for different scattered particles to describe the reflection and transmission procsesses. In effect, Eq. (\ref{eq:keup}-\ref{eq:khdn}) are utilized as wave vectors in the wave functions.

\section{Wave vectors in the SC}
Based on the BTK (Blonder-Tinkham-Klapwijk) theory \cite{btk}, the Hamiltonian for the superconductor (SC), using a field operator basis $\psi = [\psi_\uparrow, \psi_\downarrow, \psi_\uparrow^\dag, \psi_\downarrow^\dag]^T$, is given by
\begin{equation}
    \hat{H}_\text{SC}=\begin{pmatrix}
 -{\frac{\hbar^2\triangledown^2}{2 m_e}}-\mu&0&0&\Delta\\0& -{\frac{\hbar^2\triangledown^2}{2 m_e}}-\mu&-\Delta&0\\0&-\Delta^*&{\frac{\hbar^2\triangledown^2}{2 m_e}}+\mu&0\\ \Delta^*&0&0&{\frac{\hbar^2\triangledown^2}{2 m_e}}+\mu
\end{pmatrix}.
\end{equation}
The superconducting gap is denoted as $\Delta = \Delta_0 g(\theta_S)$, where $\Delta_0$ is the gap amplitude and $g(\theta_S)$ describes the superconducting pair symmetry. $\theta_S$ is the scattering angle in the SC, which can be determined from $\theta$ in the AM by using conservation of momentum along the $y$ direction. 

In a $s$-wave SC, the superconducting gap is isotropic, i.e., $g(\theta_S)=1$. The four eigenpairs are obtained as: $E_1=E_+$ with $(u_0,0,0,v_0)^T$ for $e\uparrow$, $E_2=E_-$ with $(0,u_0,-v_0,0)^T$ for $e\downarrow$, $E_3=-E_+$ with $(0,v_0,-u_0,0)^T$ for $h\uparrow$ and $E_4=-E_-$ with $(v_0,0,0,u_0)^T$ for $h\downarrow$. The eigenenergies are described by  
\begin{equation}
E_+=E_-=\sqrt{(\frac{\hbar^2(q_x^2+q_y^2)}{2m_e}-\mu)^2+|\Delta|^2}.
\label{eq:EpmS}
\end{equation}
Applying $E_1=E_2=E_3=E_4=E$, we have the usual coherence factors $u_0=\sqrt{\frac{1}{2}(1+\frac{\sqrt{E^2-\Delta_0^2}}{E})}$ and $v_0=\sqrt{\frac{1}{2}(1-\frac{\sqrt{E^2-\Delta_0^2}}{E})}$. The corresponding $x$-components of the wave vectors in the $s$-wave SC are given by
\begin{align}
q_{e}&=\frac{\sqrt{2m_e(\mu+\sqrt{E^2-\Delta_0^2})-\hbar^2q_y^2}}{\hbar},\\
q_{h}&=\frac{\sqrt{2m_e(\mu-\sqrt{E^2-\Delta_0^2})-\hbar^2q_y^2}}{\hbar},
\end{align}
which describe the electron-like and hole-like quasiparticles, respectively. Here we consider the transverse component of the wave vector is conserved across the interface, i.e., $q_y = k_y$ in the AM.

In a $d$-wave SC, the superconducting gap is anisotropic, i.e., $g(\theta_S)=\cos{(2\theta_S-2\beta)}$, in which $\beta$ defines the $d$-wave type. Due to the gap anisotropy, the eigenvectors are modified compared with those in the $s$-wave case: $(u_\pm,0,0,v_\pm e^{-i\gamma_{\pm}})^T$ for $e\uparrow$, $(0,u_\pm,-v_\pm e^{-i\gamma_{\pm}},0)^T$ for $e\downarrow$, $(0,v_\pm e^{i\gamma_{\pm}},-u_\pm,0)^T$ for $h\uparrow$ and  $(v_\pm e^{i\gamma_{\pm}},0,0,u_\pm)^T$ for $h\downarrow$. Depending on the quasiparticle motion direction, the coherence factors are $u_{\pm}=\sqrt{\frac{1}{2}(1+\frac{\sqrt{E^2-\Delta_0^2 g^2(\theta_{\pm})}}{E})}$ and $v_{\pm}=\sqrt{\frac{1}{2}(1-\frac{\sqrt{E^2-\Delta_0^2 g^2(\theta_{\pm})}}{E})}$ with $\theta_+ = \theta_S$ and $\theta_- = \pi - \theta_S$. In addition, the factor $e^{i\gamma_{\pm}} =\frac{g(\theta_{\pm})}{|g(\theta_{\pm})|}$ is introduced. Consequently, the $x$-components of the wave vectors become 
\begin{align}
q_{e,\pm}=\frac{\sqrt{2m_e(\mu+\sqrt{E^2-\Delta_0^2 g^2(\theta_{\pm})})-\hbar^2 q_y^2}}{\hbar},\\
q_{h,\pm}=\frac{\sqrt{2m_e(\mu-\sqrt{E^2-\Delta_0^2 g^2(\theta_{\pm})})-\hbar^2 q_y^2}}{\hbar},
\end{align}
where the $\pm$ sign in the subscript represents the propagation of the quasiparticles along the $\pm x$ axis. Again, $q_y = k_y$ is applied for the conservation of momentum along the $y$ direction.

Note that the energy-dependent wave-vectors and coherence factors in the SC as introduced above are only applicable for positive energies, i.e., $E>0$. When $E<0$, the following replacements should be made: $q_{e(h)} \rightarrow q_{h(e)}$, $u_0 \rightarrow -v_0^*$ and $v_0 \rightarrow u_0^*$ for $s$-wave and $q_{e(h),\pm} \rightarrow q_{h(e),\pm}$,  $u_\pm \rightarrow -v_\pm^*$ and $v_\pm \rightarrow u_\pm^*$ for $d$-wave. A detailed explanation regarding the negative energy wave vectors and coherence factors can be found in the Appendix of Ref. \cite{Sun2022Dec}.

In the following, we will focus on the $d$-wave SC since the $s$-wave case can be treated as a simplified version of $d$-wave with $g(\theta_\pm)=1$. 

\section{Wavefunctions in the AM and SC}
Aiming to investigate the differential conductance of the AM/SC bilayer system as shown in Fig. 1 in the main text, we focus on different incident particles from the AM side. 

Consider the $e\uparrow$ incident from the AM side based on the AM/SC bilayer, we have
\begin{equation}
\Psi_{\text{AM},e\uparrow}=
\begin{pmatrix}
1\\0
\end{pmatrix}e^{ik_{e\uparrow,+}x}+r
\begin{pmatrix}
 1\\0   
\end{pmatrix}e^{ik_{e\uparrow,-}x} + r_A
\begin{pmatrix}
    0\\1
\end{pmatrix}e^{ik_{h\downarrow,+}x}, 
\label{eq:AM_e_up}
\end{equation}
\begin{equation}
\Psi_{\text{SC},e\uparrow}=
t\begin{pmatrix}
u_+\\v_+ e^{-i\gamma_+}
\end{pmatrix}e^{iq_{e,+}x}+t_A
\begin{pmatrix}
 v_- e^{i \gamma_-}\\u_-  \end{pmatrix}e^{-iq_{h,-}x}, 
 \label{eq:SC_e_up}
\end{equation}
in which we use $k_y = k_{y,e\uparrow}$ given in Eq. (\ref{eq:ky_eup}). $r$, $r_A$, $t$ and $t_A$ describe the normal reflection, Andreev reflection, normal transmission and Andreev transmission, respectively, whose values can be solved by applying appropriate boundary conditions (see the next section for details).

Consider the $e\downarrow$ incident from the AM side based on the AM/SC bilayer, we have
\begin{equation}
\Psi_{\text{AM},e\downarrow}=
\begin{pmatrix}
1\\0
\end{pmatrix}e^{ik_{e\downarrow,+}x}+r
\begin{pmatrix}
 1\\0   
\end{pmatrix}e^{ik_{e\downarrow,-}x} + r_A\begin{pmatrix}
    0\\1
\end{pmatrix}e^{ik_{h\uparrow,+}x},
\end{equation}
\begin{equation}
\Psi_{\text{SC},e\downarrow}=
t\begin{pmatrix}
u_+\\-v_+ e^{-i\gamma_+}
\end{pmatrix}e^{iq_{e,+}x}+t_A
\begin{pmatrix}
 v_- e^{i \gamma_-}\\-u_-  \end{pmatrix}e^{-iq_{h,-}x},
\end{equation}
in which we use $k_y = k_{y,e\downarrow}$ given in Eq. (\ref{eq:ky_edn}).

Consider the $h\uparrow$ incident from the AM side based on the AM/SC bilayer, we have
\begin{equation}
\Psi_{\text{AM},h\uparrow}=
\begin{pmatrix}
0\\1
\end{pmatrix}e^{ik_{h\uparrow,-}x}+r
\begin{pmatrix}
 0\\1   
\end{pmatrix}e^{ik_{h\uparrow,+}x} + r_A
\begin{pmatrix}
    1\\0
\end{pmatrix}e^{ik_{e\downarrow,-}x}, 
\label{eq:AM_h_up}
\end{equation}
\begin{equation}
\Psi_{\text{SC},h\uparrow}=
t\begin{pmatrix}
v_-e^{i\gamma_-}\\-u_-
\end{pmatrix}e^{-iq_{h,-}x}+t_A
\begin{pmatrix}
 u_+ \\-v_+e^{-i \gamma_+}  \end{pmatrix}e^{iq_{e,+}x}, 
 \label{eq:SC_h_up}
\end{equation}
in which we use $k_y = k_{y,h\uparrow}$ given in Eq. (\ref{eq:ky_hup}).

Consider the $h\downarrow$ incident from the AM side based on the AM/SC bilayer, we have
\begin{equation}
\Psi_{\text{AM},h\downarrow}=
\begin{pmatrix}
0\\1
\end{pmatrix}e^{ik_{h\downarrow,-}x}+r
\begin{pmatrix}
 0\\1   
\end{pmatrix}e^{ik_{h\downarrow,+}x} + r_A
\begin{pmatrix}
    1\\0
\end{pmatrix}e^{ik_{e\uparrow,-}x}, 
\label{eq:AM_h_dn}
\end{equation}
\begin{equation}
\Psi_{\text{SC},h\downarrow}=
t\begin{pmatrix}
v_-e^{i\gamma_-}\\u_-
\end{pmatrix}e^{-iq_{h,-}x}+t_A
\begin{pmatrix}
 u_+ \\v_+e^{-i \gamma_+}  \end{pmatrix}e^{iq_{e,+}x}, 
 \label{eq:SC_h_dn}
\end{equation}
in which we use $k_y = k_{y,h\downarrow}$ given in Eq. (\ref{eq:ky_hdn}).

In the SC, the approximation $q_{e,+}\approx q_{e,-} \approx q_{h,+}\approx q_{h,-} = q_{F}$ can be applied since $E \ll \mu$ is considered in this work. Therefore, the scattering angle $\theta_S$ in the SC can be related to the incident angle $\theta$ in the AM as $\theta_S = \Real[\arctan(\frac{q_F}{ky})] $ due to the conservation of momentum along the $y$ direction. Note here the $\Real[\cdots]$ is necessary to cover the case when there exists no point on the SC Fermi surface which can satisfy conservation of $k_y$ in the AM. For the same reason, we apply real parts of all wave vectors in the SC, e.g., $q_{e,+}=\Real[q_{e,+}]$.

\section{Boundary conditions}\label{sec:BC}
To derive the boundary condition, we write down the electron Hamiltonian of the bilayer system as 
\begin{equation}
    \underline{H} =  -\frac{\hbar^2\nabla^2}{2m_e} + U_0 \delta(x) + \frac{\alpha k_y}{2} \{k_x,\Theta(-x)\} \underline{\sigma_z},
\label{eq:BC}
\end{equation}
in which only the terms affecting the boundary conditions are included, i.e., the superconducting gap terms are excluded. The anticommutator is necessary to ensure hermiticity of the Hamilton-operator and $\Theta(x)$ is the step function. Above, $k_x = -\i\partial_x$. Eq. (\ref{eq:BC}) can be rewritten as 
\begin{equation}
    {H} =  -\frac{\hbar^2\nabla^2}{2m_e} + U_0 \delta(x) + \frac{\alpha k_y \sigma}{2} \{k_x,\Theta(-x)\},
    \label{eq:BC2}
\end{equation}
where $\sigma=+1(-1)$ for $e\uparrow (\downarrow)$. In Eq. (\ref{eq:BC2}), we have
\begin{equation}
\begin{aligned}
    \{k_x,\Theta(-x)\}\Psi &= k_x[\Theta(-x)\Psi]+\Theta(-x)(k_x\Psi)\\
    &=-\i[\Psi\partial_x\Theta(-x)+\Theta(-x)\partial_x \Psi]-\i\Theta(-x)\partial_x\Psi\\
    &=\i\delta(x)\Psi-2\i\Theta(-x)\partial_x\Psi.
\end{aligned}
\end{equation}

Apply $H\Psi = E \Psi$ and integrate over $[-\epsilon,\epsilon]$ with $\epsilon \rightarrow 0$, we have
\begin{equation}
    \int_{-\epsilon}^{+\epsilon}\partial_x^2\Psi dx = \frac{2m_e}{\hbar^2}\int_{-\epsilon}^{+\epsilon}(U_0 + \frac{\i\alpha k_y\sigma}{2})\delta(x)\Psi dx - \frac{2m_e}{\hbar^2}\int_{-\epsilon}^{+\epsilon}\i\alpha k_y \sigma \Theta(-x) \partial_x \Psi dx - \frac{2m_e}{\hbar^2}\int_{-\epsilon}^{+\epsilon} E\Psi dx.
\end{equation}
Consequently, the remaining nonzero terms are
\begin{equation}
\partial_x\Psi\big|_{+\epsilon}-\partial_x\Psi\big|_{-\epsilon}=\frac{2m_e}{\hbar^2}(U_0 + \frac{\i\alpha k_y\sigma}{2})\Psi\big|_{+\epsilon}
\end{equation}
with $\Psi\big|_{+\epsilon} = \Psi\big|_{-\epsilon}$ and $\sigma=+1(-1)$ for $e\uparrow (\downarrow)$.

For notation convenience, we rewrite the boundary conditions for $e\uparrow$ incident from the AM side based on the AM/SC bilayer as
\begin{equation}
\Psi_{\text{AM}}\big|_{x=0}=\Psi_{\text{SC}}\big|_{x=0}=\begin{pmatrix}
    f\\g
\end{pmatrix},
\label{eq:BC_1}
\end{equation} 
\begin{equation}
    \partial_x\Psi_{\text{SC}}\big|_{x=0}-\partial_x\Psi_{\text{AM}}\big|_{x=0}=\begin{pmatrix}
k_{\alpha,+1}f\\k_{\alpha,-1}g
\end{pmatrix},
\label{eq:BC_2}
\end{equation} 
where $k_{\alpha,\sigma} =\frac{2m_e}{\hbar^2}(U_0 + \frac{\i\alpha k_y\sigma}{2}) $ with $\sigma=+1(-1)$.

The boundary conditions for $h\downarrow$ incident from the AM side have the same forms as Eqs. (\ref{eq:BC_1},\ref{eq:BC_2}). On the other hand, for $e\downarrow$ and $h\uparrow$ incidents from the AM side, the second boundary condition described by Eq. (\ref{eq:BC_2}) changes to
\begin{equation}
    \partial_x\Psi_{\text{SC}}\big|_{x=0}-\partial_x\Psi_{\text{AM}}\big|_{x=0}=\begin{pmatrix}
k_{\alpha,-1}f\\k_{\alpha,+1}g
\end{pmatrix},
\end{equation}

\section{DOS in the AM}
\label{sec:DOS}
For $e\uparrow$ incident from the AM side based on the AM/SC bilayer, we have 
\begin{equation}
E = E_+ =\frac{\hbar^2(k_{x}^2+k_{y}^2)}{2m_e}-\mu + \alpha k_x k_y. 
\label{eq:ellipse}
\end{equation}
The general expression for 2D density of states is given by
\begin{equation}
   N(E) = \frac{1}{4\pi^2} \int \frac{dl}{|\nabla_{\veck} E(\veck)|}, 
\label{eq:def_dos}
\end{equation}
which can be used for anisotropic DOS.

i) When $\alpha < \hbar^2/m_e$, Eq. (\ref{eq:ellipse}) defines an elliptical energy surface. In Eq. (\ref{eq:def_dos}), we can use
\begin{equation}
    dl = \sqrt{(\frac{dk_x}{d\theta})^2+(\frac{dk_y}{d\theta})^2}d\theta,
\label{eq:dl}
\end{equation}
\begin{align}
    |\nabla_{\veck} E(\veck)| &= \sqrt{(\frac{\partial E}{\partial k_x})^2 + (\frac{\partial E}{\partial k_y})^2}\notag \\
    &= \sqrt{(\frac{\hbar^2 k_x}{m_e} + \alpha k_y)^2 + (\frac{\hbar^2 k_y}{m_e} + \alpha k_x)^2}.
\label{eq:dEdk}
\end{align}
Insert $k_x = k_{x,e\uparrow}$ and $k_y = k_{y,e\uparrow}$ in Eq. (\ref{eq:ky_eup}) into Eqs. (\ref{eq:dl}) and (\ref{eq:dEdk}), $|\nabla_{\veck} E(\veck)|$ is expressed in terms of $E$ and $\theta$, i.e., $|\nabla_{\veck} E(\veck)| = K(E,\theta)$. Consequently, Eq. (\ref{eq:def_dos}) can be rewritten as 
\begin{align}
    N(E) &= \int_0^{2\pi} N(E,\theta) d\theta,\\
    N(E,\theta) &= \frac{1}{4\pi^2} \frac{\sqrt{(dk_{x,e\uparrow}/d\theta)^2+(dk_{y,e\uparrow}/d\theta)^2}}{K(E,\theta)}
\label{eq:DOS_AM_eup_+}
\end{align}
in which $N(E,\theta)$ corresponds to the DOS at a given incident angle $\theta$. 

ii) When $\alpha > \hbar^2/m_e$, Eq. (\ref{eq:ellipse}) corresponds to a hyperbola, which can not define a closed integral path. Therefore, we confine $\alpha < \hbar^2/m_e$ in this work, as mentioned before. 

Following the same procedure as described above, the DOS in the AM for $e\downarrow$, $h\uparrow$ and $h\downarrow$ incidents can be calculated.

\section{Conductance}
The quantum mechanical charge current density for $e\uparrow$ channel in the AM is given by
\begin{equation}
 j_{Q,e\uparrow} = -\frac{e\hbar}{m_e}[\Imag{\{f^*\nabla f \}} + \Imag{\{g^*\nabla g \}}] - \frac{e\alpha k_y}{\hbar}(|f|^2-|g|^2).
 \label{eq:j_Q,eup}
\end{equation}
The charge current density for $h\downarrow$ channel in the AM has the same form as Eq. (\ref{eq:j_Q,eup}). On the other hand, for $e\downarrow$ and $h\uparrow$ channels, the charge current density expression changes to
\begin{equation}
j_{Q,e\downarrow(h\uparrow)} = -\frac{e\hbar}{m_e}[\Imag{\{f^*\nabla f \}} + \Imag{\{g^*\nabla g \}}] + \frac{e\alpha k_y}{\hbar}(|f|^2-|g|^2).
\end{equation}

Use Eq. (\ref{eq:j_Q,eup}), we can compute the charge current contributions 1-5 to the $e\uparrow$ channel in the AM. Assume that a voltage is applied across the AM/SC junction so that distribution function for electrons is $f(E-eV)$ on the AM side while it is $f(E)$ on the SC side. 

\begin{itemize}
\item 1: contribution from the incoming $e\uparrow$ on the AM side:
\begin{align}
    \psi_1 &= \begin{pmatrix}
1\\0
\end{pmatrix}e^{ik_{e\uparrow,+}x},\\
     j_{Q1} &= -e(\frac{\hbar k_{e\uparrow,+}}{m_e}+\frac{\alpha k_y}{\hbar}).
\end{align}
This contributes to the total charge current density in the $e\uparrow$ channel on the AM side with 
\begin{align}
    J_{Q1} &= f(E-eV)j_{Q1}\notag\\
    &=-ef(E-eV)(\frac{\hbar k_{e\uparrow,+}}{m_e}+\frac{\alpha k_y}{\hbar}).
\end{align}

\item 2: contribution from the reflected $e\uparrow$ produced by the incoming $e\uparrow$ on the AM side:
\begin{align}
    \psi_2 &= r_1\begin{pmatrix}
1\\0
\end{pmatrix}e^{ik_{e\uparrow,-}x},\\
     j_{Q2} &= -e|r_1|^2(\frac{\hbar k_{e\uparrow,-}}{m_e}+\frac{\alpha k_y}{\hbar}),
\end{align}
in which $r_1$ is the solved reflection coefficient from the wave functions given by Eqs. (\ref{eq:AM_e_up},\ref{eq:SC_e_up}). 
This contributes to the total charge current density in the $e\uparrow$ channel on the AM side with 
\begin{align}
    J_{Q2} &=f(E-eV)j_{Q2}\notag\\
    &=-e|r_1|^2f(E-eV)(\frac{\hbar k_{e\uparrow,-}}{m_e}+\frac{\alpha k_y}{\hbar}).
\end{align}

\item 3: contribution from the transmitted $e\uparrow$ produced by the incoming $e\uparrow$ on the SC side:
\begin{align}
    \psi_3 &= t_2\begin{pmatrix}
1\\0
\end{pmatrix}e^{ik_{e\uparrow,-}x},\\
     j_{Q3} &= -e|t_2|^2(\frac{\hbar k_{e\uparrow,-}}{m_e}+\frac{\alpha k_y}{\hbar}),
\end{align}
in which $t_2$ is the solved transmission coefficient from the following wave functions:
\begin{equation}
\Psi_{\text{SC}}= \begin{pmatrix}
u_-\\v_- e^{-i\gamma_-}
\end{pmatrix}e^{-iq_{e,-}x} + r_2 \begin{pmatrix}
u_+\\v_+ e^{-i\gamma_+}
\end{pmatrix}e^{iq_{e,+}x}+r_{A2}
\begin{pmatrix}
 v_- e^{i \gamma_-}\\u_-  \end{pmatrix}e^{-iq_{h,-}x},   
\end{equation}
\begin{equation}
\Psi_{\text{AM}}= t_2
\begin{pmatrix}
1\\0
\end{pmatrix}e^{ik_{e\uparrow,-}x} + t_{A2}
\begin{pmatrix}
 0\\1   
\end{pmatrix}e^{ik_{h\downarrow,+}x}.
\end{equation}
This contributes to the total charge current density in the $e\uparrow$ channel on the AM side with 
\begin{align}
    J_{Q3} &=f(E)j_{Q3}\notag\\
    &=-e|t_2|^2f(E)(\frac{\hbar k_{e\uparrow,-}}{m_e}+\frac{\alpha k_y}{\hbar}).
\end{align}

\item 4: contribution from the Andreev-reflected $e\uparrow$ produced by the incoming $h\downarrow$ on the AM side:
\begin{align}
    \psi_4 &= r_{A3}\begin{pmatrix}
1\\0
\end{pmatrix}e^{ik_{e\uparrow,-}x},\\
     j_{Q4} &= -e|r_{A3}|^2(\frac{\hbar k_{e\uparrow,-}}{m_e}+\frac{\alpha k_y}{\hbar}),
\end{align}
in which $r_{A3}$ is the solved Andreev reflection coefficient from the wave functions given by Eqs. (\ref{eq:AM_h_dn},\ref{eq:SC_h_dn}). This contributes to the total charge current density in the $e\uparrow$ channel on the AM side with 
\begin{align}
    J_{Q4} &=[1-f(-E-eV)]j_{Q4}\notag\\
    &=-e|r_{A3}|^2[1-f(-E-eV)](\frac{\hbar k_{e\uparrow,-}}{m_e}+\frac{\alpha k_y}{\hbar}).
\end{align}

\item 5: contribution from the Andreev-transmitted $e\uparrow$ produced by the incoming $h\downarrow$ on the SC side:
\begin{align}
    \psi_5 &= t_{A4}\begin{pmatrix}
1\\0
\end{pmatrix}e^{ik_{e\uparrow,-}x},\\
     j_{Q5} &= -e|t_{A4}|^2(\frac{\hbar k_{e\uparrow,-}}{m_e}+\frac{\alpha k_y}{\hbar}),
\end{align}
in which $t_{A4}$ is the solved Andreev transmission coefficient from the following wave functions: 
\begin{equation}
\Psi_{\text{SC}}= \begin{pmatrix}
v_+ e^{i\gamma_+}\\u_+
\end{pmatrix}e^{iq_{h,+}x}+
r_4\begin{pmatrix}
v_- e^{i\gamma_-}\\u_-
\end{pmatrix}e^{-iq_{h,-}x}+r_{A4}
\begin{pmatrix}
 u_+\\v_+ e^{-i \gamma_+}\end{pmatrix}e^{iq_{e,+}x},
\end{equation}
\begin{equation}
\Psi_{\text{AM}}=t_4
\begin{pmatrix}
0\\1
\end{pmatrix}e^{ik_{h\downarrow,+}x}+t_{A4}
\begin{pmatrix}
 1\\0   
\end{pmatrix}e^{ik_{e\uparrow,-}x}.
\end{equation}

This contributes to the total charge current density in the $e\uparrow$ channel on the AM side with 
\begin{align}
    J_{Q5} &=f(E)j_{Q5}\notag\\
    &=-e|t_{A4}|^2f(E)(\frac{\hbar k_{e\uparrow,-}}{m_e}+\frac{\alpha k_y}{\hbar}),
\end{align}
\end{itemize}

When computing the differential conductance as described in the main text, only contributions induced by incident particles from the AM side contribute since we have chosen to apply the bias voltage there, e.g., the differential conductance originates from $j_{Q3}$ and $j_{Q5}$ becomes zero when calculating $dI/dV$ in the $e\uparrow$ channel. As a result, we only need to consider incidents from the AM side, as mentioned before. The physics is unchanged if one chooses to apply the voltage in a different manner, so long as the voltage difference between the AM and SC is the same.

\section{45 degree rotated AM}
Here we summarize the useful equations for the rotated Hamiltonian, i.e.,
\begin{equation}
    \underline{H}_\text{AM} =  -{\frac{\hbar^2\triangledown^2}{2 m_e}} - \mu + \frac{\alpha}{2}(k_x^2- k_y^2)\underline{\sigma_z},
\end{equation}
which corresponds to a 45 degree
rotation of the AM/SC interface. 
\begin{itemize}
\item 1: eigenpairs:
 
 The four eigenpairs are obtained as: $E_1=E_+$ with $(1,0,0,0)^T$ for $e\uparrow$, $E_2=E_-$ with $(0,1,0,0)^T$ for $e\downarrow$, $E_3=-E_+$ with $(0,0,1,0)^T$ for $h\uparrow$ and $E_4=-E_-$ with $(0,0,0,1)^T$ for $h\downarrow$. The eigen-energies are described by  
\begin{equation}
E_\pm=\frac{\hbar^2(k_{x}^2+k_{y}^2)}{2m_e}-\mu \pm \frac{\alpha}{2} (k_x^2-k_y^2).
\end{equation} 

\item 2: wave vectors in the AM to construct the wave functions:
\begin{align}
k_{e\uparrow,\pm}&=\pm \sqrt{\frac{2m_e(\mu+E+\alpha k_y^2/2)-\hbar^2k_y^2}{\hbar^2+m_e \alpha}},\\
k_{e\downarrow,\pm}&=\pm \sqrt{\frac{2m_e(\mu+E-\alpha k_y^2/2)-\hbar^2k_y^2}{\hbar^2-m_e \alpha}},\\
k_{h\uparrow,\pm}&=\pm \sqrt{\frac{2m_e(\mu-E+\alpha k_y^2/2)-\hbar^2k_y^2}{\hbar^2+m_e \alpha}},\\
k_{h\downarrow,\pm}&=\pm \sqrt{\frac{2m_e(\mu-E-\alpha k_y^2/2)-\hbar^2k_y^2}{\hbar^2-m_e \alpha}}.
\end{align}

\item 3: wave vectors on the AM Fermi surface:
\begin{align}
k_{y,e\uparrow} &=  r_1 \sin\theta
,\quad k_{x,e\uparrow} = r_1 \cos\theta,\notag \\r_1 &= \frac{a_1 b_1}{\sqrt{b_1^2 \cos^2(\theta+\pi/2)+a_1^2 \sin^2(\theta+\pi/2)}},\quad a_1 = \sqrt{\frac{2m_e(\mu+E)}{\hbar^2-m_e \alpha}},\quad b_1= \sqrt{\frac{2m_e(\mu+E)}{\hbar^2+m_e\alpha}}\\
k_{y,e\downarrow} &= r_2 \sin\theta
,\quad k_{x,e\downarrow} = r_2 \cos\theta,\notag \\r_2 &= \frac{a_2 b_2}{\sqrt{b_2^2 \cos^2\theta+a_2^2 \sin^2\theta}},\quad a_2 = \sqrt{\frac{2m_e(\mu+E)}{\hbar^2-m_e \alpha}},\quad b_2= \sqrt{\frac{2m_e(\mu+E)}{\hbar^2+m_e\alpha}}\\
k_{y,h\uparrow} &=  r_3 \sin\theta
,\quad k_{x,h\uparrow} = r_3 \cos\theta,\notag \\r_3 &= \frac{a_3 b_3}{\sqrt{b_3^2 \cos^2(\theta+\pi/2)+a_3^2 \sin^2(\theta+\pi/2)}},\quad a_3 = \sqrt{\frac{2m_e(\mu-E)}{\hbar^2-m_e \alpha}},\quad b_3= \sqrt{\frac{2m_e(\mu-E)}{\hbar^2+m_e\alpha}}\\
k_{y,h\downarrow} &= r_4  \sin\theta
,\quad k_{x,h\downarrow} = r_4 \cos\theta,\notag \\r_4 &= \frac{a_4 b_4}{\sqrt{b_4^2 \cos^2\theta+a_4^2 \sin^2\theta}},\quad a_4 = \sqrt{\frac{2m_e(\mu-E)}{\hbar^2-m_e \alpha}},\quad b_4= \sqrt{\frac{2m_e(\mu-E)}{\hbar^2+m_e\alpha}}
\end{align}


\item 4: boundary conditions:
\begin{equation}
\Psi_{\text{AM}}\big|_{x=0}=\Psi_{\text{SC}}\big|_{x=0}=\begin{pmatrix}
    f\\g
\end{pmatrix},
\end{equation} 
\begin{equation}
    \partial_x\Psi_{\text{SC}}\big|_{x=0}-\begin{pmatrix}
(1+m_e\alpha/\hbar^2)\partial_x f\\(1-m_e\alpha/\hbar^2)\partial_x g
\end{pmatrix}\big|_{x=0}=\frac{2m_eU_0}{\hbar^2}\begin{pmatrix}
    f\\g
\end{pmatrix}
\end{equation} 
for $e\uparrow$ and $h\downarrow$ incidents. To get the above boundary conditions, we follow the similar procedure as described in Appendix \ref{sec:BC} by considering the Hermitian electron Hamiltonian of the bilayer system as 
\begin{equation}
    \underline{H} =  -\frac{\hbar^2\nabla^2}{2m_e} + U_0 \delta(x) + \frac{\alpha}{2}[k_x \Theta(-x) k_x - k_y \Theta(-x) k_y]\underline{\sigma_z},
\end{equation}
in which $k_x = -\i\partial_x$.

On the other hand, we have
\begin{equation}
    \partial_x\Psi_{\text{SC}}\big|_{x=0}-\begin{pmatrix}
(1-m_e\alpha/\hbar^2)\partial_x f\\(1+m_e\alpha/\hbar^2)\partial_x g
\end{pmatrix}\big|_{x=0}=\frac{2m_eU_0}{\hbar^2}\begin{pmatrix}
    f\\g
\end{pmatrix}
\end{equation}
for $e\downarrow$ and $h\uparrow$ incidents.

\item 5: charge current density expressions for different channels:
\begin{align}
j_{Q,e\uparrow(h\downarrow)} &= -\frac{e\hbar}{m_e}[\Imag{\{f^*\nabla f \}} + \Imag{\{g^*\nabla g \}}] - \frac{e\alpha}{\hbar}[\Imag{\{f^*\nabla f \}} - \Imag{\{g^*\nabla g \}}],\\
j_{Q,e\downarrow(h\uparrow)} &= -\frac{e\hbar}{m_e}[\Imag{\{f^*\nabla f \}} + \Imag{\{g^*\nabla g \}}] + \frac{e\alpha}{\hbar}[\Imag{\{f^*\nabla f \}} - \Imag{\{g^*\nabla g \}}].
\end{align}

\end{itemize}

\section{Arbitrary-angle rotated AM}
The arbitrary-angle rotated AM can be modeled based on the combination of our established 0 and 45 degree cases, i.e., a more general Hamiltonian is
\begin{equation}
    \underline{H}_\text{AM} =  -{\frac{\hbar^2\triangledown^2}{2 m_e}} - \mu + \alpha_1 k_x k_y\underline{\sigma_z} + \alpha_2 (k_x^2- k_y^2)\underline{\sigma_z}/2,
\end{equation}
in which two different altermagnetism strength parameters $\alpha_1$ and $\alpha_2$ are introduced and the arbitrary angle is determined by $\theta_\alpha = \frac{1}{2}\arctan(\alpha_1/\alpha_2)$. Following the same procedure as introduced before, the eigenvalues and wave vectors can be solved from the Hamiltonian, e.g.,
\begin{equation}
E_\pm=\frac{\hbar^2(k_{x}^2+k_{y}^2)}{2m_e}-\mu \pm \alpha_1 k_xk_y\pm \frac{\alpha_2}{2} (k_x^2-k_y^2),
\end{equation}
\begin{align}
    k_{e\uparrow,\pm}=&\pm\frac{1}{\hbar+
    \alpha_2m_e/\hbar}\sqrt{2m_e(\mu+E)(1+\frac{\alpha_2m_e}{\hbar^2})-\hbar^2 k_y^2 +\frac{(\alpha_1^2+\alpha_2^2) m_e^2 k_y^2}{\hbar^2}}\notag\\
    &-\frac{\alpha_1 m_e k_y}{\hbar^2+m_e\alpha_2},
\end{align}
which reveal features of both the 0 and 45 degree cases. To ensure that the energy dispersion corresponds to an elliptical energy surface rather than a hyperbola, the altermagnetism parameters should satisfy $\bar \alpha \equiv \sqrt{\alpha_1^2+\alpha_2^2} < \alpha_c \equiv \hbar^2/m_e$. The corresponding semi-major and semi-minor axes are $a=\sqrt{\frac{2m_e(\mu+E)}{\hbar^2-m_e\bar\alpha }} $ and $b=\sqrt{\frac{2m_e(\mu+E)}{\hbar^2+m_e\bar\alpha }} $ for electron incidents, based on which the DOS can be calculated. Similarly, the boundary conditions and charge currents expressions can be derived from the Hamiltonian with all necessary details included in our previous explanation for the 0 and 45 degree cases.

\section{Ferromagnet}

To model a normal ferromagnet (FM), we use the Hamiltonian
\begin{equation}
    \underline{H}_\text{FM} =  -{\frac{\hbar^2\triangledown^2}{2 m_e}} - \mu + J_\text{ex}\underline{\sigma_z},
\end{equation}
in which $J_\text{ex}$ is the exchange energy in the FM.
\begin{itemize}
\item 1: eigenpairs:
 The four eigenpairs are obtained as: $E_1=E_+$ with $(1,0,0,0)^T$ for $e\uparrow$, $E_2=E_-$ with $(0,1,0,0)^T$ for $e\downarrow$, $E_3=-E_+$ with $(0,0,1,0)^T$ for $h\uparrow$ and $E_4=-E_-$ with $(0,0,0,1)^T$ for $h\downarrow$. The eigen-energies are described by  
\begin{equation}
E_\pm=\frac{\hbar^2(k_{x}^2+k_{y}^2)}{2m_e}-\mu \pm J_\text{ex}.
\end{equation} 

\item 2: wave vectors in the FM to construct the wave functions:
\begin{align}
k_{e\uparrow,\pm}&=\pm \frac{1}{\hbar} \sqrt{2m_e(\mu+E-J_\text{ex})-\hbar^2k_y^2},\\
k_{e\downarrow,\pm}&=\pm \frac{1}{\hbar} \sqrt{2m_e(\mu+E+J_\text{ex})-\hbar^2k_y^2},\\
k_{h\uparrow,\pm}&=\pm \frac{1}{\hbar} \sqrt{2m_e(\mu-E-J_\text{ex})-\hbar^2k_y^2},\\
k_{h\downarrow,\pm}&=\pm \frac{1}{\hbar} \sqrt{2m_e(\mu-E+J_\text{ex})-\hbar^2k_y^2}.
\end{align}

\item 3: wave vectors on the FM Fermi surface:
\begin{align}
k_{y,e\uparrow} &=  r_1 \sin\theta
,\quad k_{x,e\uparrow} = r_1 \cos\theta,\quad r_1 = \frac{1}{\hbar}\sqrt{2m_e(\mu+E-J_\text{ex})}\\
k_{y,e\downarrow} &= r_2 \sin\theta
,\quad k_{x,e\downarrow} = r_2 \cos\theta, \quad r_2 = \frac{1}{\hbar}\sqrt{2m_e(\mu+E+J_\text{ex})}\\
k_{y,h\uparrow} &=  r_3 \sin\theta
,\quad k_{x,h\uparrow} = r_3 \cos\theta,\quad r_3 = \frac{1}{\hbar}\sqrt{2m_e(\mu-E-J_\text{ex})}\\
k_{y,h\downarrow} &= r_4  \sin\theta
,\quad k_{x,h\downarrow} = r_4 \cos\theta,\quad r_4 = \frac{1}{\hbar}\sqrt{2m_e(\mu-E+J_\text{ex})}.
\end{align}
Following the same approach as described for AM, the DOS for FM can be derived based on the above new wavevectors. Here we found the DOS at angle $\theta$ is given by
\begin{equation}
    N(E,\theta) = \frac{m_e}{4\pi^2\hbar^2},
\end{equation}
which is the same for $e\uparrow$, $e\downarrow$, $h\uparrow$ and $h\downarrow$ incidents.

\item 4: boundary conditions:
\begin{equation}
\Psi_{\text{FM}}\big|_{x=0}=\Psi_{\text{SC}}\big|_{x=0}=\begin{pmatrix}
    f\\g
\end{pmatrix},
\end{equation} 
\begin{equation}
    \partial_x\Psi_{\text{SC}}\big|_{x=0}-\partial_x\Psi_{\text{FM}}\big|_{x=0} = \frac{2m_eU_0}{\hbar^2}\begin{pmatrix}
    f\\g
\end{pmatrix}.
\end{equation} 
These boundary conditions apply for $e\uparrow$, $e\downarrow$, $h\uparrow$ and $h\downarrow$ incident cases.

\item 5: charge current density expressions for different channels:
\begin{equation}
j_{Q} = -\frac{e\hbar}{m_e}[\Imag{\{f^*\nabla f \}} + \Imag{\{g^*\nabla g \}}]
\end{equation}
which has the same form for $e\uparrow$, $e\downarrow$, $h\uparrow$ and $h\downarrow$ channels.
\end{itemize}

Based on the above expressions, we can investigate the charge and spin conductances for the FM/SC bilayer. We have done so and our results agree with \cite{FMSC}, in which the charge conductance decreases with increasing $J_\text{ex}$.

\section{Andreev-reflection probability}
We here determine the Andreev reflection probabilities for different incident angles.

The probability coefficients are derived by applying the continuity of the probability current at the $k_x^2-k_y^2$ AM/SC interface. In the $k_x^2-k_y^2$ AM, if we write the wave function in the form of $\Psi_\text{AM} = (f,g)^T$, the probability current is given by  
\begin{equation}
j_P^{\text{AM}} = \frac{\hbar}{m_e} [\Imag{\{f^*\nabla f \}} - \Imag{\{g^*\nabla g \}}]\pm^{'}\frac{\alpha}{\hbar}[\Imag{\{f^*\nabla f \}} + \Imag{\{g^*\nabla g \}}],
\end{equation}
in which $\pm^{'}=+$ for $e\uparrow(h\downarrow)$ incident and $\pm^{'}=-$ for $e\downarrow(h\uparrow)$ incident. In the SC, if we write the wave function in the form of $\Psi_\text{SC} = (f,g)^T$, the probability current is given by

\begin{equation}
    j_P^{\text{SC}} = \frac{\hbar}{m_e} [\Imag{\{f^*\nabla f \}} - \Imag{\{g^*\nabla g \}}],
\end{equation}
which has the same form for $e\uparrow$, $e\downarrow$, $h\uparrow$ and $\downarrow$ incidents. By applying $j_P^{\text{SC}} \big|_{x=0} = j_P^{\text{AM}} \big|_{x=0}$ and inserting the explicit expressions of the wavefunctions, the probability coefficients of the Andreev reflection, normal reflection, Andreev transmission and normal transmission can be derived, and the sum of the four probability coefficients induced by the same incident is as 1.  

\begin{figure}
\includegraphics[width=0.99\columnwidth]{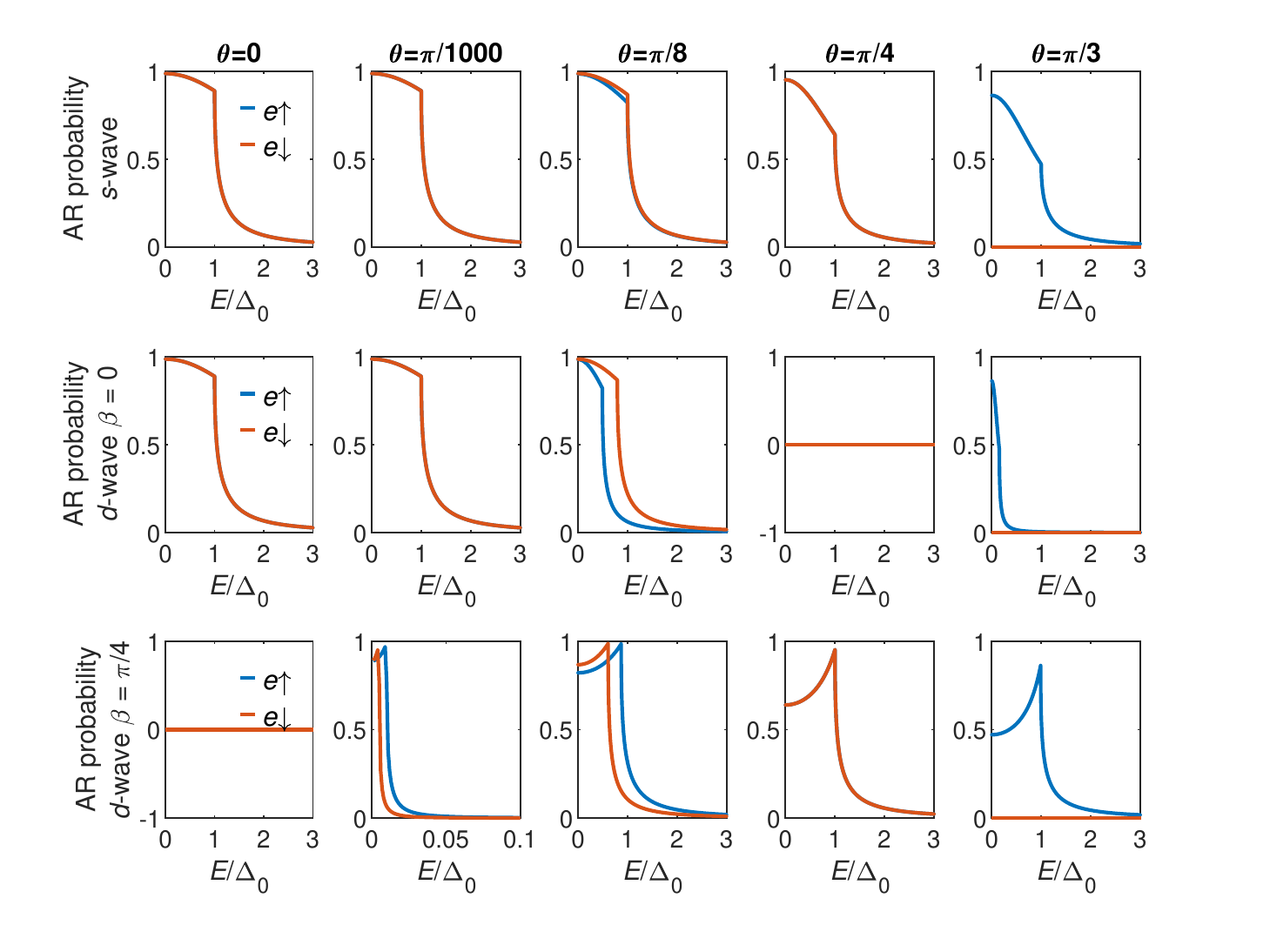}
	\caption{(Color online) Andreev reflection (AR) probability for different spin species [$e\uparrow(\downarrow)$] channels as a function of energy for different types of $k_x^2-k_y^2$ AM/SC junctions and incident angles $\theta$: the rows correspond to different superconducting order parameter symmetries and the columns correspond to different $\theta$ . Here we use $Z=0$ and $\alpha/\alpha_c=0.6$
	}
	\label{AR_PROB}
\end{figure}

Except for the AR probability, the normal reflection (NR) probability should also be considered since NR suppresses the conductance. We here focus on a particular example: the $d$-wave $\beta=\pi/4$ SC at a small incident angle, e.g., $\theta = \pi/8$. Unlike the AR, in order to get conductance in the $e\uparrow(\downarrow)$ channel through NR, the NR probability is derived based on the wavefunctions induced by the $e\uparrow(\downarrow)$ incident. In addition, we compare the probability behaviors between AM/SC and FM/SC, as shown in the Fig. \ref{A_and_B}. 

\begin{figure}
\includegraphics[width=0.99\columnwidth]{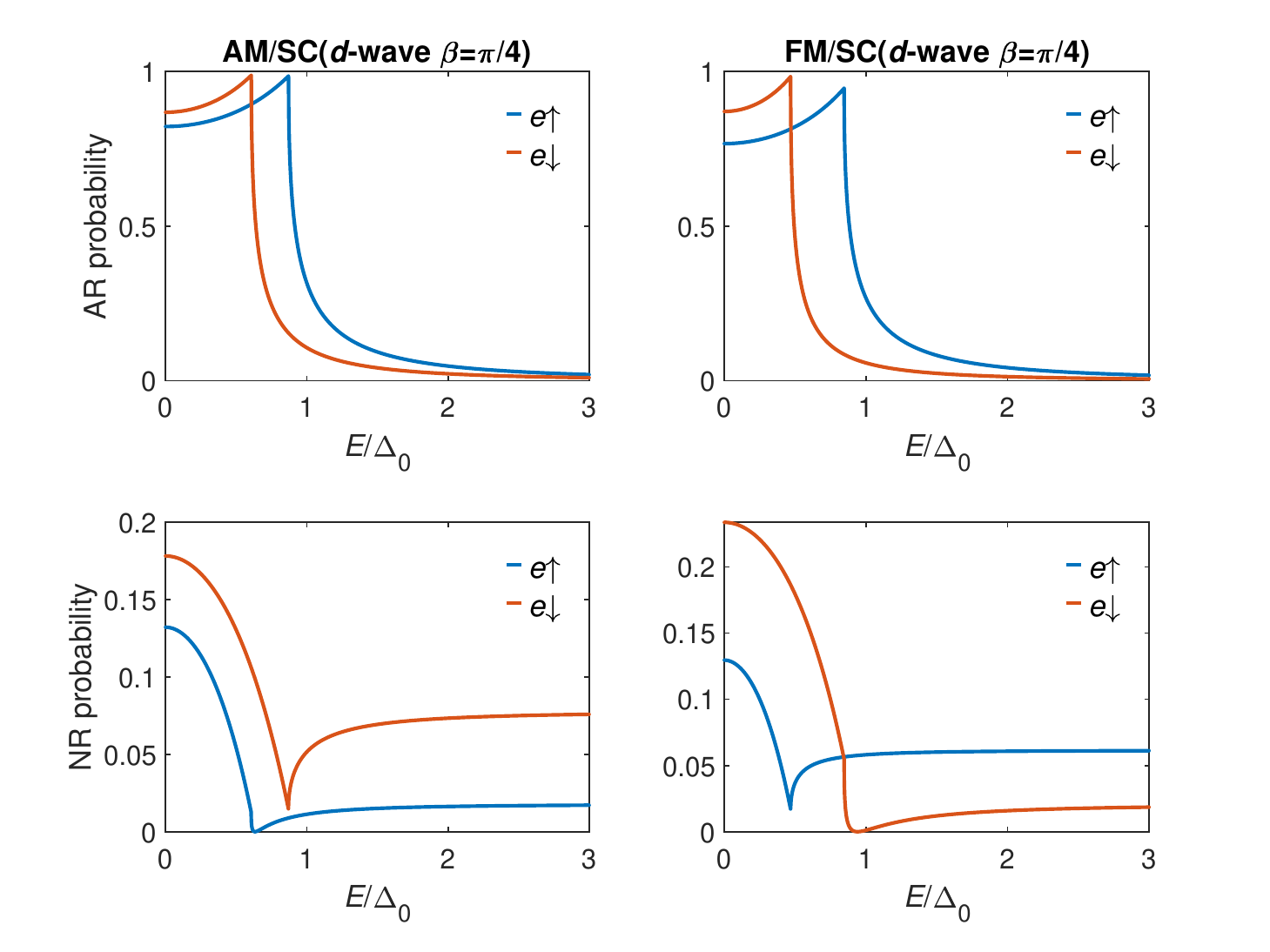}
	\caption{(Color online) Probability coefficients for different spin species [$e\uparrow(\downarrow)$] channels as a function of energy for different junctions at a small incident angle $\theta = \pi/8$ : the rows correspond to different reflection probabilities and the columns correspond to different junctions. Here we use $Z=0$, $\alpha/\alpha_c=0.6$ for AM and $J_\text{ex}/\mu=0.6$ for FM.
	}
	\label{A_and_B}
\end{figure}


\end{widetext}

\end{document}